\documentclass[12pt,preprint]{aastex}
\usepackage[usenames,dvips]{color}
\def\msun{{\rm ~M}_{\odot}}
\def\ergs{{\rm\,erg\,s^{-1}}}

\def\be{\begin{equation}}
\def\ee{\end{equation}}
\begin{document}

\title{Accretion Disk Spectra of the Brightest Ultra-luminous X-ray Source 
in M82}

\author{
Feng Yuan\altaffilmark{1,}\altaffilmark{2}
Ronald E. Taam\altaffilmark{3,}\altaffilmark{4},\\
R. Misra\altaffilmark{5},
Xue-Bing Wu\altaffilmark{6},
Yongquan Xue\altaffilmark{7}}

\altaffiltext{1}{Shanghai Astronomical Observatory, 80 Nandan Road, Shanghai 200030, China}
\altaffiltext{2}{Joint Institute for Galaxy and Cosmology (JOINGC) of SHAO
and USTC}
\altaffiltext{3}{Northwestern University, Department of Physics and Astronomy, 2145 Sheridan Road, Evanston, IL 60208}
\altaffiltext{4}{ASIAA/National Tsing Hua University - TIARA, Hsinchu, Taiwan}
\altaffiltext{5}{Inter-University Center for Astronomy and Astrophyscs, Post Bag 4, 
Ganeshkhind, Pune 411 007, India}
\altaffiltext{6}{Department of Astronomy, Peking University, Beijing 100871, China}
\altaffiltext{7}{Department of Physics, Purdue University, West Lafayette, IN 47907}

\begin{abstract}
Emission spectra of hot accretion disks characteristic of advection dominated 
accretion flow (ADAF) models are investigated for comparison with the 
brightest ultra-luminous source, X-1, in the galaxy M82. If the spectral state 
of the source is similar to the low luminosity hard state of stellar mass black 
holes in our Galaxy, a fit to the {\it Chandra} X-ray spectrum and constraints 
from the radio and infrared upper limits, require a black hole mass in the range 
of $9 \times 10^4 - 5 \times 10^5 \msun$. Lower black hole masses ($\la 10^4 \msun$) 
are possible if M82 X-1 corresponds to the high luminosity hard state of Galactic 
black hole X-ray binary sources. Both of these spectrally degenerate hot accretion 
disk solutions lead to an intermediate mass black hole interpretation for M82 X-1. 
Since these solutions have different spectral variability with X-ray luminosity and
predict different infrared emission, they can be distinguished by future off axis
{\it Chandra} observations or simultaneous sensitive infrared detections. 
\end{abstract}

\keywords{black hole physics - galaxies: individual: M82: - X-rays: 
galaxies - X-rays: black holes}

\section{Introduction}

Ever since the discovery of ultra-luminous X-ray sources (ULXs) in external 
galaxies with the {\it Einstein} satellite (see Fabbiano 1989), much attention 
has focused on their nature as they represent a class of abnormally bright 
X-ray sources ($\sim 10^{39} - 10^{41} \ergs$). The existence of 
temporal variability on a variety of timescales in this subpopulation 
suggests that these systems are accreting objects.  For recent reviews, 
see Miller \& Colbert (2004) and Miller (2005). Within this framework 
ULXs have been considered as either stellar mass black holes accreting at 
super Eddington rates or intermediate mass black holes (IMBHs) accreting at 
sub Eddington rates (see Colbert \& Mushotzky 1999; Makishima et al. 2000).  
In the case of stellar mass black holes, super Eddington luminosities (by a 
factor of 10) can be produced from optically thick, slim accretion disks 
(Watarai, Mizuno \& Mineshige 2001). Alternatively, apparent super Eddington 
luminosities can result in circumstances where significant beaming of radiation 
can occur (King et al. 2001; Begelman 2002; K\"ording, Falcke, \& Markoff 2002) 
with enhancement factors ranging from less than 5 for funnel shaped disks (see 
Misra \& Sriram 2003) to $\sim 100$ for a beamed relativistic jet (see K\"ording 
et al 2002). In both descriptions, the effect is directionality dependent and is 
maximal for face on viewing.  We note that the theoretical models which invoke 
beaming do not address the {\it spectrum} of ULXs as the primary focus is toward 
providing an understanding of the {\it luminosity}. In the IMBH interpretation, the 
challenge is to provide an understanding for the formation of such massive black holes. 
 
In addition to their high X-ray luminosities, hints of their nature should also be 
revealed by their emission spectra.  Spectral transitions, similar to those seen 
in Galactic black hole systems, have been reported in  NGC 1313 X-1 (Colbert \& 
Mushotzky 1999) and two sources in IC342 (Kubota et al. 2001). Analysis of 
{\it XMM-Newton} data revealed the presence of a cool accretion disk component 
($kT_{in} \sim 0.1-0.5 $ keV) in NGC 1313 X-1, X-2 (Miller et al. 2003) and M81 X-9 
(Miller, Fabian, \& Miller 2004), which suggested that these sources harbor IMBH. 
However, Gon{\c c}alves \& Soria (2006) have argued that such soft spectral components 
depend on the complexity of the fitting model. Indeed, the spectra of ULXs have been 
fitted by a variety of empirical and physical models that include an absorbed power 
law, a multi-temperature disk blackbody, spectral cutoffs, and/or a combination of the 
above.  While an absorbed power law can adequately represent most (but not all) of 
the spectra observed by {\it Chandra}, spectral fits of bright ULXs observed by 
{\it XMM Newton} often require a spectral break or cutoff above 2 keV (Stobbart, 
Roberts, \& Wilms 2006).  In general, the spectra of ULXs reveal a diversity of 
spectral shapes, suggesting that the sources identified as ULXs are a heterogeneous 
class and/or exhibit various spectral states. Thus, a detailed spectral study of a 
single source may offer better evidence on the presence (or absence) of IMBHs.

In attempting to distinguish between the accreting stellar mass and IMBH models, detailed 
calculations of the emission spectrum from an accretion disk surrounding the black hole 
are required.  Specifically, the fitting of both the luminosity and spectrum of a 
source is necessary to quantitatively confront the theoretical models with observations.
Among the ULXs, the bright ULX source X-1 in M82 is a very promising candidate for
this purpose because its luminosity has been measured to be
as high as 1.6 $\times$ 10$^{41}\ergs$ (Ptak \& Griffiths 1999). 
The early {\it XMM-Newton} spectral (Fiorito \& Titarchuk 2004; Agrawal \& 
Misra 2006) and temporal observational results (Strohmayer \& Mushotzky 2003) 
may not be conclusive because recent {\it Chandra} observations (taken 
on 4-5 Feb 2005) by Kaaret, Simet 
\& Lang (2006) revealed that there are two bright nearby X-ray sources which 
would be contributing to the {\it XMM-Newton} flux.
We note that the observation of Kaaret et al. (2006) was taken such that
M82 X-1 was off axis and, hence, the data was not affected by pile-up effects. This
allowed, for the first time, a non ambiguous (i.e. with possibly little contamination)
measure of the source spectral properties. The spectrum of the brightest X-ray
source, X41.4+60, is well described by a power law
with photon index of $\Gamma = 1.67 \pm 0.02$ with an isotropic luminosity in the
2-10 keV energy range of $2.4 \times  10^{40}\ergs $. Hereafter we will use the 
name M82 X-1 to denote X41.4+60.

It is reasonable to hypothesize that the hydrodynamic and radiative models invoked for 
explaining the spectra of stellar mass black holes could also be applicable to systems 
with IMBHs. It is well known that observationally stellar mass black hole systems 
exhibit different spectral states, denoted as quiescent, hard/low, intermediate, soft/high, 
and very high (or steep power law) states (see McClintock \& Remillard 2006 for a review).
These states are distinguished by their spectral and timing properties. 
For example, in a typical hard state, the
X-ray spectrum is well described by a power law form with $\Gamma=1.5 \sim 2.1$
and the radio emission is relatively strong. 
On the other hand, the X-ray spectrum is much steeper
in the soft state with $\Gamma=2.1 \sim 4.8$ with no detectable radio emission. With
respect to their timing properties, the integrated power spectrum is usually strong in
the hard state and often exhibits a QPO. This is in contrast to the timing properties
in the soft state where the power spectrum is very weak and QPOs are not observed.
It is possible that the spectral states of ULXs can be similarly described.
This idea has been invoked in the study of ULXs by Winter, Mushotzky, \& Reynolds 
(2006), and what is more, similar state transitions to those in stellar mass
black holes have been detected for two ULXs in NGC 1313 (Feng \& Kaaret 2006).

The spectrum of M82 X-1 unambiguously resembles the low/hard state of Galactic 
black hole X-ray binary systems, which is much harder than that of the very high
state. Therefore, it is possible that M82 X-1 can be explained in terms of
emission from an advection dominated accretion flow (ADAF) since such accretion flows 
have been successfully applied to understand the properties of the hard state of
these binary systems 
(McClintock \& Remillard 2006). Thus, this source is an ideal candidate to 
investigate whether, indeed, these systems can also be described by ADAF solutions and 
to test the hypothesis that they harbor IMBHs. Moreover, there are radio and 
infrared observations of this region which, as is described in a later section, place 
further constraints and consistency checks on these models.
  
In this paper, we focus on the calculation of the theoretical spectrum of M82 X-1 and 
present a detailed emission spectrum from an hot accretion disk model as applied to 
ULXs.  A number of calculations have been carried out for a range of black hole masses, 
mass accretion rates, and other input parameters with the goal of fitting the observed 
X-ray spectrum as well as the isotropic X-ray luminosity (since beaming is unimportant in 
such models) to provide constraints on the properties of the system.  In the next section, 
we provide a description of the accretion disk model, which forms the basis of our 
study, and the parameters used to fit the spectrum.  The numerical results of the 
detailed emission spectrum are presented for a range of black hole masses and mass flow 
rates in the disk and compared to the observed spectrum of M82 X-1 in \S 3.  Finally, 
we discuss the implications of our results and conclude in the last section. 

\section{Model Description}

The origin of the X-ray emission characterizing the hard state is very likely due to 
the Comptonization process of soft photons by thermal electrons in a hot accretion disk 
corona (see e.g., Zdziarski 2000 and Zdziarski \& Gierli\'{n}ski 2004 for reviews). Such 
a corona may be produced during the process of magnetic reconnection above the standard 
thin disk (Liang \& Price 1977; Galeev, Rosner, \& Vaiana 1979). However, this process 
is poorly understood and detailed models have yet to be developed. Moreover, Esin et al. 
(2001) show that in the case of XTE J1118+480 the EUV data requires that the standard thin 
disk must be truncated at a certain radius. On the other hand, a hot corona is a natural 
consequence of an ADAF (Narayan, Mahadevan \& Quataert 1998), which is dynamically well 
described and, hence, its astrophysical applications can be examined in detail. 

In this paper, we concentrate on the ADAF model. Narayan (1996) and Narayan, McClintock, 
\& Yi (1996) first proposed to apply the ADAF to the hard state of stellar mass black 
holes. Subsequently, Esin, McClintock \& Narayan (1997) presented detailed calculations 
applied to Nova Muscae 1991. Yuan, Cui \& Narayan (2005) further developed the Esin et 
al. (1997) model by taking into account the presence of outflows in ADAFs (Blandford 
\& Begelman 1999) and by including a jet component which appears to be always present 
in the hard state (Fender 2006).

Briefly, the hard spectral state can be described in terms of an accretion flow 
consisting of two components, namely a standard thin disk external to a transition 
radius, $r_{\rm tr}$, and an ADAF (or luminous hot accretion flow; see Yuan et al. 
2006 for details) interior to it. We assume that only a fraction of the mass flow 
rate at $r_{\rm tr}$ accretes onto the black hole with the remainder ejected in the 
form of an outflow.  The strength of this outflow is described by a parameter $s_0$.  
The possible existence of this outflowing material was hypothesized in analytical 
work (Narayan \& Yi 1994; Blandford \& Begelman 1999) and has been confirmed in 
numerical simulations (Stone, Pringle, \& Begelman 1999; Hawley \& Balbus 2002; 
Igumenshchev, Narayan, \& Abramowicz 2003).  Following Yuan et 
al. (2005), we assume that $\dot M(r)=\dot M_0$ for $r\geq r_{\rm tr}$, and
\be\frac{{\rm d} \ln \dot{M}(r)}{{\rm d} \ln r} = s(r), \qquad r< r_{\rm tr}
\label{mdot}\ee where
\be s(r) = s_0 \max[f(r),0],\ee
and $s_0$ is independent of $r$. Here, the advection term $f(r)$ is defined as the 
ratio of the rates of energy advection to viscous heating.  When $\dot M$ is very 
low, $f(r)=1$ and $s(r)=s_0$. In this case, eq.\ (\ref{mdot}) gives us the usual
form, $\dot{M} = \dot{M}_0 (r/r_{\rm tr})^{s_0}$ (e.g., Blandford \& Begelman 
1999), where $\dot{M}_0$ is the accretion rate at $r_{\rm tr}$.  In the innermost 
regions, some fraction of the inflowing material is assumed to be redirected to flow 
in the vertical direction as a jet.  We denote the mass loss rate in the jet as 
$\dot{M}_{\rm jet}$. 

The X-ray emission in the hard state, in this model, originates 
from the thermal Comptonization of seed photons associated with synchrotron emission 
in the ADAF. We note that there are other possible sources of seed photons, such as 
the soft photons from the standard thin disk exterior to $r_{\rm tr}$ or possibly 
the existence of cold clumps within the hot accretion flow, which may be important in 
some cases (see \S 3, 4). The temperature of the thin disk for $r > r_{\rm tr}$ is 
determined by the local viscous dissipation and the non local irradiation from the 
inner ADAF. The optical radiation and infrared/radio emission in the hard state arise 
mainly from this thin disk and the jet respectively. 

The determination of the emission spectrum from the accretion disk model requires 
specifying a number of 
input parameters.  In addition to the mass of the black hole, $M$, and mass accretion 
rate, $\dot{M}_0$, the viscous parameter $\alpha$, the ``magnetic'' parameter, $\beta$ 
(describing the strength of the magnetic field), the fraction of viscous dissipation 
that directly heats electrons, $\delta$, the transition radius $r_{\rm tr}$, and 
$s_0$ must be chosen.  Among these parameters, the values of $\alpha$, $\beta$, and 
$\delta$ are determined by the microphysical  processes associated with the 
magnetohydrodynamic (MHD) driven turbulence in the disk. We assume that these parameters 
do not differ appreciably among different sources, including 
supermassive black hole sources, Galactic X-ray binaries, ULXs and M82 X-1 
in particular. The MHD numerical simulations of 
accretion flows and previous ADAF modeling of several well-studied sources constrain 
their values to a narrow range (e.g., Hawley \& Krolik 2001; Yuan, Quataert \& 
Narayan 2003).  Based on extensive calculations covering a wide range
of values, the spectral results are insensitive to the particular choices. 
Hence, we fix their values as $\alpha=0.3, \beta=0.9,$ and $\delta=0.5$. 
The value of the transition radius, $r_{\rm tr}$, is unimportant for 
the determination of the X-ray spectrum and mainly affects the optical/UV 
spectrum where, currently, data is lacking. Therefore, its value is  
unimportant for our study, and we set $r_{\rm tr}=100 r_s$ where 
$r_s \equiv 2GM/c^2$.  The most important parameters affecting our 
modeling are the remaining three parameters, i.e., $M$, $\dot{M}_0$ 
and $s_0$.  The values of $M$ and $\dot{M}_0$ are completely unknown 
and obtaining estimates for their values is our primary goal. Although 
a value of the remaining parameter, $s_0$, has been inferred from the 
ADAF modeling of sources such as Sgr A$^*$ and XTE J1118+480 (Yuan, 
Quatatert \& Narayan 2003; Yuan, Cui \& Narayan 2005), we do not restrict 
$s_0$, instead allowing it to vary in a range as wide as possible. Thus, in 
order to obtain a satisfactory fit to the X-ray spectrum, these 
three parameters are varied.

In order to explain the radio emission (see below), an additional 
jet component is required in addition to the accretion flow, as in the case of
Galactic black hole X-ray binaries (e.g., Yuan, Cui \& Narayan 2005; Fender 2006).
We use the internal shock scenario, which is widely adopted in the study of 
gamma ray burst afterglows, to calculate the jet emission (see Yuan, Cui \& 
Narayan 2005 for details). 
Briefly, internal shocks within the jet occur due to collisions of shells with 
different velocities. The shocks accelerate a fraction of the electrons into a 
power-law energy distribution. The energy of the electrons and the strength
of the magnetic field after the shock is determined by two parameters, 
$\epsilon_{\rm e}$ and $\epsilon_B$, which give the fraction of the shock energy 
transferred into the accelerated electrons and the (amplified) magnetic field, 
respectively. The radiative transfer in the jet is calculated to obtain the
emitted spectrum. We would like to emphasize, however, that large uncertainties exist 
in the jet model and its solution is not unique. Fortunately the jet emission is
usually negligible in the X-ray band (ref. Fig. 1) and thus will not affect 
the result of our paper.

\section{Emission Spectrum of M82 X-1}

\subsection{Observed Spectrum}

The X-ray data of Kaaret et al. (2006) is described by a power law spectrum 
with photon index of $\Gamma=1.67$ and a 2-10 keV isotropic luminosity of $2.4
\times 10^{40} \ergs$.  In addition to the data in the X-ray regime, we also 
include data in the radio and infrared wavelengths in Fig. 1. 
Kaaret et al. (2006) conducted four VLA observation at a frequency of
8.5GHz between 29 Jan 2005 and 5 Feb 2005. We choose the radio 
data taken on 5 Feb 2005 since it is the closest in time (same day) to
the {\em Chandra} observation.  The radio position is offset by about 1\arcsec 
relative to the position of the brightest X-ray source (X41.4+60) in M82 
and only a marginal detection of a radio flux of 0.5$\pm$0.1 mJy has been 
obtained. However, Kaaret et al. (2006) argue that due to the uncertainty 
of the {\it Chandra} X-ray position, the radio and X-ray sources are very likely 
the same.  The corresponding isotropic radio luminosity was found to be 
$(6.7\pm1.3) \times 10^{34}$ erg s$^{-1}$ for an assumed distance of 3.63 Mpc.
An upper limit for the infrared luminosity for M82 X-1 is also shown based 
on near infrared observations of the associated star cluster MGG-11 
with the 10m Keck II telescope on 23 Feb 2002 (McCrady, Gilbert, 
\& Graham 2003). A magnitude of 13.10$\pm$0.15 at 1.6$\micron$ and 
of 12.03$\pm$0.05 at 2.2$\micron$ was determined. 
By dereddening the spectrum due to extinction, McCrady et al. 
(2003) estimate a light to mass ratio $L/M$ to be 3.5$^{+2.1}_{-1.5}$ 
(in unit of $L_\odot/M_\odot$) for MGG-11 at 1.6$\micron$. Estimating 
the mass of MGG-11 to be $(3.5 \pm 0.1) \times 10^5 \msun$, they find 
an infrared luminosity of MGG-11 at 1.6$\micron$ as $4.7^{+2.9}_{-2.2} 
\times 10^{39}$ erg s$^{-1}$. It should be noted that the infrared observation 
was not taken simultaneously with the radio and X-ray observations. Since 
it is difficult to estimate the respective contributions of the stars and of 
M82 X-1 to the IR flux, we take this value as an upper limit 
to the luminosity at 1.6$\micron$ for M82 X-1. 

\subsection{Calculated Spectrum}

From a number of calculations, the numerical results indicate that
to fit the X-ray spectrum, including the luminosity and spectral slope, the 
mass of the black hole in M82 X-1 (and correspondingly the mass accretion rate)
 must lie in a narrow range.  For $s_0 = 
0.27$, the most favored value in the case of Sgr A$^*$ (Yuan, Quataert 
\& Narayan 2003), the mass of the black hole in M82 X-1 is estimated to 
be $1.8 \times 10^5 \msun$ and the corresponding mass accretion rate 
to be $\dot{M}_0 = 0.55 \dot{M}_{\rm Edd}$ where $\dot{M}_{\rm Edd}\equiv
L_{\rm Edd}/c^2$ is the Eddington accretion rate \footnote{Here, the efficiency for 
the conversion of rest mass energy to radiation is $L_{\rm ADAF}/\dot{M}_0
=0.009L_{\rm Edd}/0.55\dot{M}_{\rm Edd} c^2=0.016$. This value is much smaller
than the efficiency of a standard thin disk and is a characteristic feature
of an ADAF. The produced spectrum is shown in Fig. 1 by the thin solid line}. 

The constraint on the mass of the black hole (and the mass accretion rate) 
can be obtained from fitting the X-ray spectrum as follows. Noting that the X-ray 
spectrum of M82 X-1 is produced by the Comptonization process in the ADAF,
the spectral slope is determined by the Compton $y$ parameter, which is 
proportional to the product of the Thompson optical depth 
and the electron temperature of the accretion flow, with a larger $y$ 
corresponding to a harder spectrum. As the mass of the black hole is increased (from 
the favored value of $1.8 \times 10^5\msun$), a smaller
$\dot{M}$ would be required to produce the same luminosity. In this case, 
the Compton $y$ parameter will become smaller due to the decrease of the density of the
accretion flow. Therefore, the predicted
spectrum would be too soft, as shown by the dashed line in Fig. 1.
On the other hand, if the mass were smaller, the predicted spectrum would
be too hard, as shown by the dot-dashed line in Fig. 1.
We note that the results also depend on the value of $s_0$. This is because
the produced X-ray spectrum is the sum of the local Comptonization spectrum from 
different radii of the ADAF which have different slope, while the value of $s_0$ 
controls the fractional contribution of each radius thus affecting the total 
spectral slope.  For $s_0$ varying from $s_0=0$ to $s_0=0.4$, a satisfactory fit 
to the X-ray spectrum can be obtained leading to black hole masses ranging from 
$M=9\times 10^4\msun$ to $M=5\times 10^5\msun$.

In Fig. 1 we also show by the dotted and dot-dot-dashed lines the predicted 
emission produced by the truncated thin disk and jet.  In this case, the mass 
loss rate in the jet is $\dot{M}_{\rm jet}=1.5 \times 10^{-3}\dot{M}_{\rm Edd}$, 
which is $\sim 0.5\%$ of the accretion rate at $5r_s$. The power
of this jet is $\sim 10^{41}\ergs$, which is comparable to the total X-ray luminosity
emitted from the ADAF. Furthermore, the two 
parameters, $\epsilon_e$ and $\epsilon_B$, were found 
to be $\epsilon_e=0.06$ and $\epsilon_B=0.02$.  All these parameter values are surprisingly 
similar to the modeling results of the hard states of the two stellar mass 
black hole sources, XTE J1118+480 and XTE J1550--564 where $\dot{M}_{\rm jet}=
0.5\%, 0.6 \%\dot{M}(5r_s)$ and the values of $\epsilon_e$ and 
$\epsilon_B$ are the same (Yuan, Cui \& Narayan 2005; Yuan et al. 2006). 
Such a similarity provides some support to our assumption that
M82 X-1 corresponds to the hard state of stellar mass black hole sources.
For the truncated thin disk, which contributes from the ultraviolet to the 
infrared portion of the spectrum, $\dot{M}=0.55\dot{M}_{\rm Edd}$ and the viewing 
angle (as defined from the disk rotation axis) is assumed to be $\theta=60^{\fdg}$. 
The parameters for the predicted spectrum yields an infrared luminosity 
of $\sim 5\times 10^{39}\ergs$. This value lies above the lower bound of the 
upper limit for the observed luminosity (see \S 3.1) as deduced for the 
lowest extinction, corresponding to $2.5 \times 10^{39}\ergs$.
Thus, this computational model requires that the extinction cannot be very 
low or the viewing angle of the thin disk is larger than $\sim 70^{\fdg}$. 

In our procedure, the mass of the black hole based on the ADAF model is obtained 
simply from the observed luminosity and the spectral slope.  This is a direct 
result of the fact that for a ``pure'' ADAF model, where the synchrotron photons
are the only seed photons, a monotonic correlation is predicted between the spectral 
slope and the Eddington-scaled luminosity.  Specifically, the ADAF models
predict that for X-ray luminosities in the range of $\sim 10^{-4}L_{\rm Edd}$
and $\sim 4 \% L_{\rm Edd}$ the spectrum hardens with increasing 
luminosity (Esin et al. 1997).  

The predicted correlation is partially confirmed by observations, as
shown in Fig. 2. Here, a correlation is found to exist between the photon 
index $\Gamma$ and $L_x/L_{\rm Edd}$ for the two stellar mass black hole
X-ray sources which exhibit the widest range in hard state luminosity. The 
data in this figure has been taken from Tomsick, Corbel \& Kaaret (2001) for 
the 2000 outburst of XTE J1550--564 and analyzed by ourselves for 
XTE J1118+480 from archival RXTE data. 
Note that we simply calculate $L_x$ by assuming that the power law spectrum 
cuts off at the same energy as given by the thin solid line in Fig. 1.
We can see from the figure that the qualitative behavior of the
correlation below $\sim 2\%L_{\rm Edd}$ is consistent with the prediction 
of the spectra produced from ADAF models.  However, the observational 
results reveal that the correlation is non monotonic. That is, 
above $\sim 2\%L_{\rm Edd}$, the spectrum softens with increasing 
luminosity.  The physical mechanism underlying this change in correlation 
is not understood.  Since the hard states approach a very high state where 
$L_x \ga 2\%L_{\rm Edd}$, it is very likely that the non monotonic behavior 
is related to an additional physical description of accretion and emission 
at the very high state.  Although this regime is not understood, it is 
possible that the accretion flow responsible for the very high state and 
relatively luminous hard state consists of two phases, with cold clumps 
embedded within hot gas. In this model, the emission of the clumps supplies
additional seed photons to those produced by the synchrotron process in the 
hot phases to harden the spectrum as a result of the Comptonization process 
in the accretion flow (see Yuan et al. 2006). This could 
result in a different correlation in comparison to the case of a ``pure'' 
one-phase ADAF model. Alternatively, this state may reflect the contribution 
of another component in the system (e.g., a consequence of the emission 
associated with a hot disk cooled by the cold disk photons). In this 
case the variation of the spectral index with luminosity could be 
distinct from the pure ADAF solution.  As the accretion rate increases, 
the transition radius decreases (e.g., Liu et al. 1999),  
the flux of seed photons and the thus the cooling of ADAF are 
stronger, reducing the Compton $y$ factor, thereby leading to softer 
spectra (e.g., Zdziarski, Lubinski, \& Smith 1999).  This would be 
in contrast to the pure ADAF model prediction, but in accordance with 
the variation observed in XTE J1550--564 at high luminosities.

The correlation displayed in Fig. 2 must be confirmed over a much larger 
sample of black hole binary X-ray sources to determine its generality.  
Existing work appears to partially confirm the existence of such a 
correlation (e.g., Shemmer et al. 2006; Remillard \& McClintock 2006), 
suggesting that the correlation 
is, indeed, multi-valued.  That is, a given spectral index, $\Gamma$, may 
correspond to two different luminosity levels with the luminosity 
difference greater for sources characterized by softer spectra.  Thus, 
it is possible that M82 X-1 may correspond to
the very luminous hard state of stellar mass black hole sources. In 
the model studied here, the ``pure'' ADAF model can only describe the
relatively dim hard state. Unfortunately, the physical description of the 
high luminosity hard state has not been developed sufficiently to present a 
detailed numerical model.  Instead, we roughly estimate the properties of a 
black hole X-ray binary source in this regime. 
From Fig. 2, and given the possible scattering in the relation due
to, for example, hysteresis effects (see Zdziarski \& Gierli\'{n}ski 2004),
a spectrum with $\Gamma \simeq 1.67$ can be reached at luminosities as 
high as $L\la 4-10\%
L_{\rm Edd}$.  In this case, the cut-off energy of the X-ray luminosity should 
be slightly lower than in the ADAF model and, thus, the bolometric luminosity 
of M82 X-1 should be slightly lower, $\sim 10^{41}\ergs$. The mass of the 
black hole in M82 X-1 would then be in the range of $\sim (1-2)\times 
10^4 \msun$ and the required mass accretion rate may be as high as 
$\dot{M}_{\rm Edd}$ according to our modeling experience in XTE J1550--564 
(Yuan et al. 2006). We note that for the high luminosity model our calculation
shows that the predicted 
luminosity at $\sim 10^{14}$ Hz is about an order of magnitude lower 
than the model illustrated in Fig. 1, thus satisfying the observed infrared upper 
limit by a larger margin.
 
\section{Implications}

In this paper, we have presented a detailed theoretical emission spectrum 
for M82 X-1, within the advection dominated accretion flow framework, which 
is widely used to model the relatively low luminosity hard state of Galactic 
black hole X-ray binaries. The results of our detailed numerical calculations 
reveal that a X-ray photon power law spectrum of index $\Gamma = 1.67$ can be 
reproduced by the synchrotron self Comptonization process of the thermal electrons 
in the ADAF. The X-ray luminosity in this solution is $L_{\rm x} \approx 
0.9\%L_{\rm Edd}$ (note that it is $L_{\rm x}$, not $L_{\rm 2-10 keV}$). 
 A comparison of the model results with the constraints 
imposed by the X-ray spectral index and luminosity yield a mass for the black 
hole in the range from $9 \times 10^4 - 5\times 10^5 \msun$.  The model 
parameters required are similar to those obtained by detailed fitting of 
stellar mass black holes, with the exception of the high value of the black 
hole mass. 

The accretion disk modeling of the system, however, is not unique since 
observations of stellar mass black hole sources, such as XTE J1550--564,
reveal the existence of a luminous hard state. These states can have the same 
spectral slope of $\Gamma = 1.67$, but at luminosities that are significantly 
higher, $L_{\rm x} \approx 4-10\% L_{\rm Edd}$. For this regime the accretion 
efficiency is likely higher and can be comparable to the efficiencies 
characteristic of a standard optically thick disk (see Yuan et al. 2006). 
The detailed accretion flow model for these luminous hard states is, however, 
still lacking, but the cause for the multi-valued nature of the accretion 
flow spectral solutions is likely associated with the presence of additional 
seed photons for the Comptonization in the hot accretion flow. This may 
arise from the outer cool disk or from the cool clumps within a highly 
inhomogeneous accretion flow structure, perhaps, analogous to the two phase
disks discussed by Yuan et al. (2006) and Merloni, Malzac, Fabian, \& Ross (2006).  
If M82 X-1 is in such a luminous hard state, the mass of the black hole could be 
reduced by an order of magnitude to $\sim 10^4 \msun$. 

To discriminate the possible solution at high luminosity from the ``pure'' 
ADAF one at low luminosity, future simultaneous observations will be necessary. 
A particularly important wavelength region for study is the infrared where 
the infrared emission associated with the high luminosity solution is significantly 
lower than at low luminosities. A better observational estimate for the infrared 
luminosity would entail a more precise determination of the extinction and 
removal of the infrared contribution from the stellar cluster MGG-11.  If the 
obtained infrared luminosity were to lie below the prediction of the thick solid 
line in Fig. 1, the model with $M=1.8 \times 10^5 \msun$ 
can be ruled out.  We note that such a solution would alleviate the severe 
constraints on the required viewing angle of the system if the observed 
infrared luminosity is significantly less than our adopted upper limit.
An additional discriminant between the two models can be 
explored in the hard X-ray band, where the model shown in Fig. 1 predicts a 
cut-off energy of $\sim 100$ KeV.  The cut-off energy for the high luminosity, 
low black hole mass model should be lower, perhaps $\sim 30-50$ keV.  
Hence, X-ray observations in the hard X-ray band  
can also be used to discriminate between these models. 

Our modeling results suggest that an ADAF interpretation for ULXs is viable 
and supports the idea that M82 X-1 can harbor an IMBH with a mass close to the 
AGN limit of $10^5-10^6 \msun$ (Greene \& Ho 2004; Barth, Greene, \& Ho 2005). 
Our results on the low luminosity state of black hole X-ray binaries as applied 
to M82 X-1, however, are in conflict with the results based on the hypothesis 
that it is a binary system with an orbital period of 62 days.  Such a period 
was inferred from an {\it RXTE} X-ray monitoring campaign of M82 by Kaaret, 
Simet, \& Lang (2006).  If this variability is attributed to M82 X-1 orbiting
about its common center of mass in a binary system, evolutionary 
calculations for the system place constraints on the masses of the binary 
system components. In particular, Patruno et al. (2006) find a solution in 
which a donor in the mass range of $22-25 \msun$ transfers mass to an 
IMBH of mass $200 - 5000 \msun$ under the assumption that the system is 
associated with the star cluster MGG-11.  This latter assumption constrains 
the age of the system to be in the range of 7-12 Myr. Although a solution 
was also found for a stellar mass black hole accreting at super Eddington 
rates, this solution was considered unlikely due to its very restrictive 
parameter space.  Very recently, Okajima, Ebisawa, \& Kawaguchi
(2006) have proposed a stellar mass black hole interpretation for M82 X-1
based on {\it XMM Newton} observational data.  In their model, a slim disk
description has been adopted. However, the existence of multiple sources in the
{\it XMM Newton} field of view of M82 X-1 makes their interpretation premature
for the brightest source, X41.4+60.

We point out that the inferred black hole masses are dependent on the 
radiative efficiency of the accretion process.  For example, the estimates 
based on binary evolution theory implicitly assume an efficiency of $\sim 
0.1$, whereas the efficiencies can range from 0.01 to $\sim 0.1$ in the hot 
accretion disk solutions for the low luminosity and high luminosity regimes 
respectively. If one adopts the estimated mass in the high luminosity solution 
of $10^4\msun$, then it is only a factor of 2 higher than the estimated upper 
limit of $5000\msun$ from the binary evolutionary calculations. There are 
uncertainties in the estimate of the black hole mass in the high luminosity 
solution, as well as in the study by Patruno et al. (2006).  Such uncertainties 
arise, on the one hand, because the orbital period in the mass transferring 
state is rather insensitive for black holes of $M_{\rm bh} \ga 1000 \msun$ and, 
on the other, the radiative efficiencies for hot accretion flows in the high 
luminosity are not well determined.  Thus, taking into account accretion disk 
theory as well as binary evolutionary theory, and given the uncertainties, a 
mass for the black hole in M82 X-1 may well lie in the range of $5000 - 10^4 
\msun$.

Future X-ray observations of M82 X-1 will be essential to determine the 
viability of hot accretion flows to this ULX and its nature.  In particular, 
it would be of particular importance to determine if the correlation between 
the X-ray luminosity and the photon spectral index is confirmed as M82 X-1 
varies. Such a test would necessarily involve future {\it Chandra} off axis 
observations since {\it XMM} studies of this region are affected by source 
confusion and previous {\it Chandra} on axis observations suffer from pile up 
effects.  Thus, the results of this study highlight the need for additional 
future off axis {\it Chandra} observations of this source.  Studies of M82 
X-1 at energies greater than 10 keV are also important since curvature in 
the X-ray spectrum is expected from the thermal Comptonization of electrons 
in the coronae of standard optically thick disks or slim disk models, whereas
a power law spectrum is expected for hot disk models.  Thus, such observations 
have the potential to discriminate between the optically thick slim disk 
models from the optically thin hot accretion disk models. 


\acknowledgements

We are grateful for useful discussions with Dr. Albert Kong on the X-ray data 
and interpretation. We thank the anonymous referee for the many 
constructive suggestions. This work was supported, in part, by the Theoretical 
Institute for Advanced Research in Astrophysics (TIARA) operated under 
Academia Sinica and the National Science Council Excellence Projects program 
in Taiwan administered through grant number NSC 95-2752-M-007-006-PAE.  FY is 
partially supported by the One-Hundred-Talent Program of China and Pujiang Program.

{}

\clearpage

\begin{figure} \epsscale{1.} \plotone{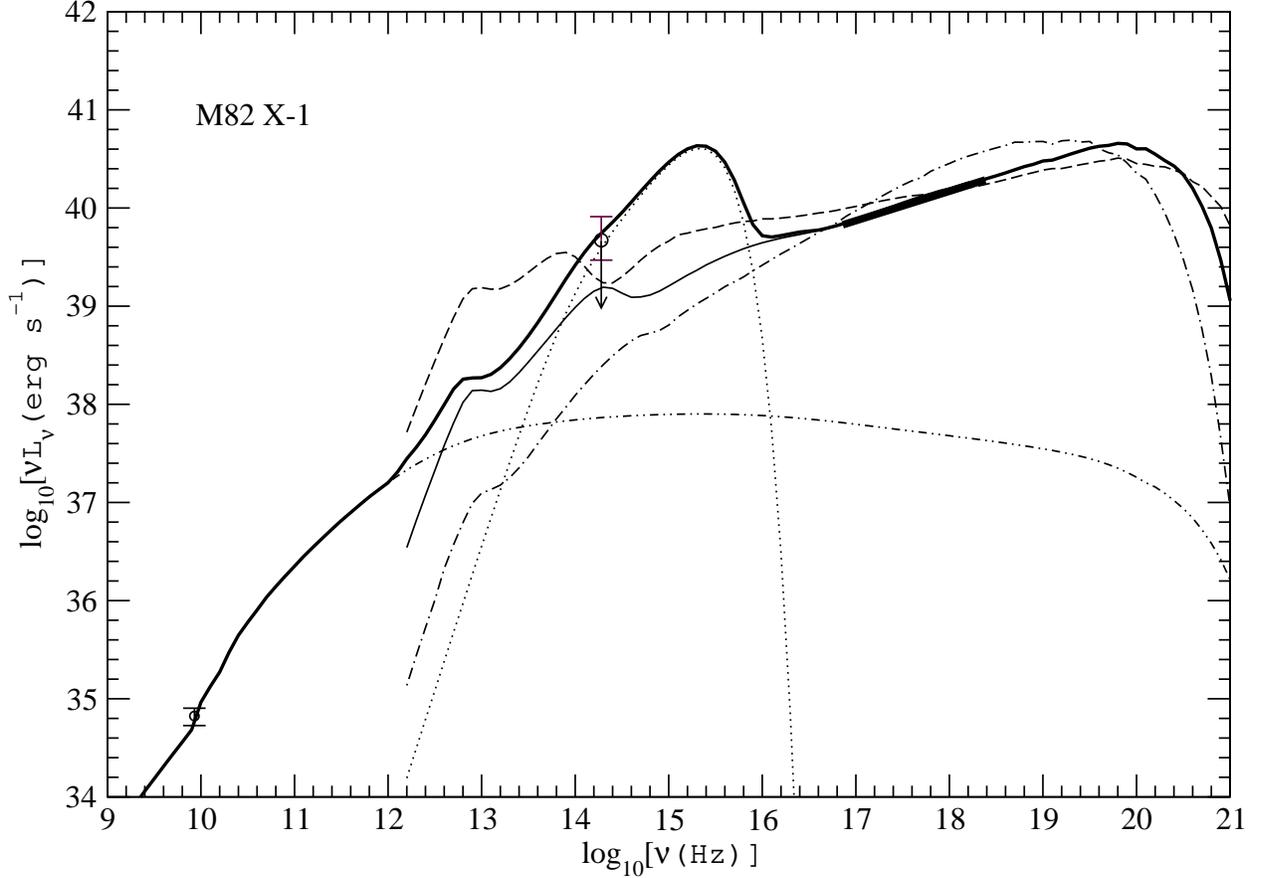} \vspace{0.5in}
\caption{ADAF model for M82 X-1. The open circle with arrow and error bars should 
be regarded as the upper limit, with the error bar mainly coming from the uncertainty
of the extinction for the infrared data point. The X-ray data is
denoted by the thick line segment.  The thin solid line shows the spectrum produced
by the ADAF with $M=1.8 \times 10^5 \msun$, $\dot{M} (R_{\rm tr})=
0.55\dot{M}_{\rm Edd}$. The dotted and dot-dot-dashed lines are for the
emitted spectra from the thin disk truncated at $R_{\rm tr}=100R_s$ and
a jet. Their sum is shown by the thick solid line. 
The dashed and dot-dashed lines are produced by two ADAFs with 
heavier and lighter black hole masses, respectively. They are 
obviously too soft and hard compared to the observed spectrum. 
See text for details.}
\end{figure}

\begin{figure} \epsscale{1.} \plotone{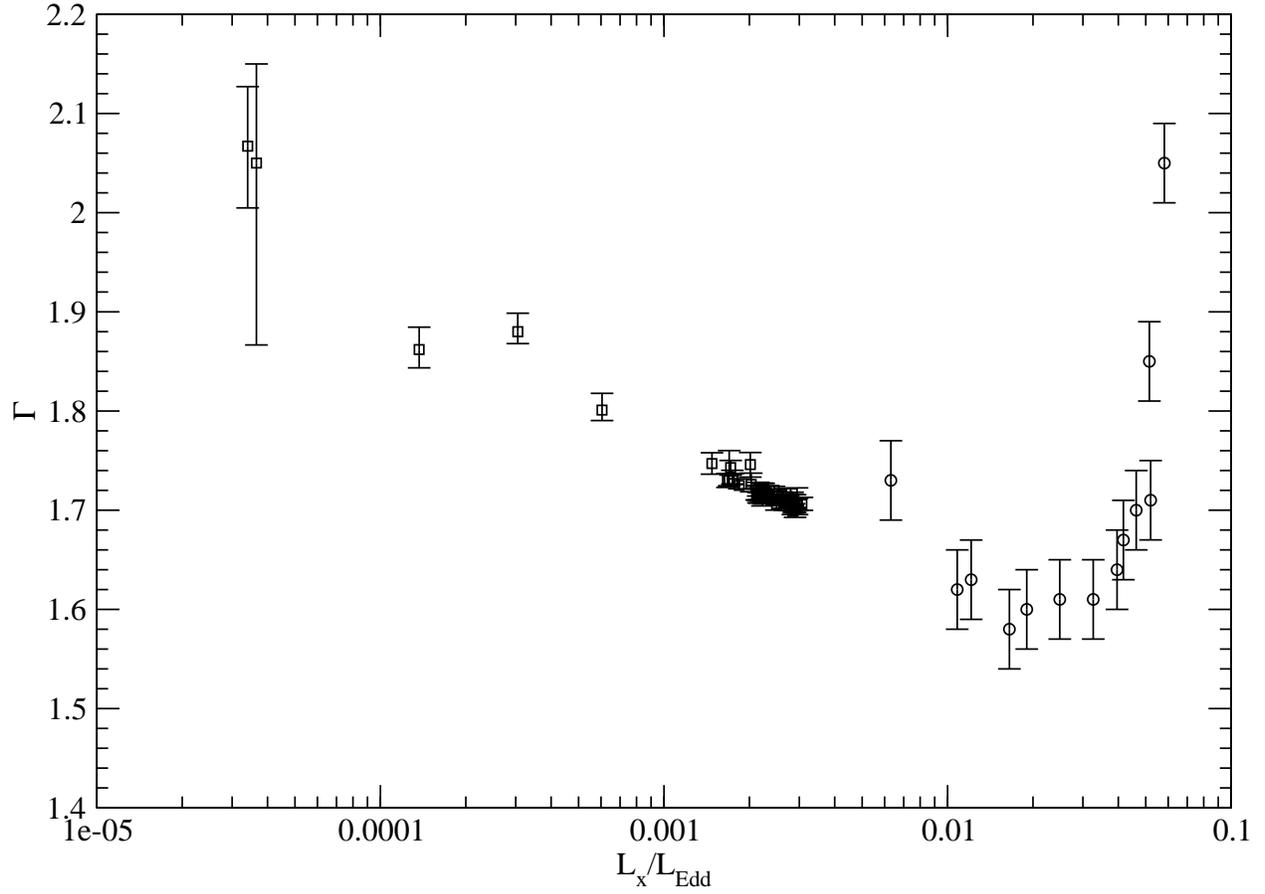} \vspace{1.in}
\caption{The correlation between the photon index and the X-ray luminosity for
two stellar mass black hole sources, XTE J1118+480 (open squares) 
and XTE J1550--564 (open circles).}
\end{figure}
\end{document}